\documentclass[journal=jacsat,manuscript=article]{achemso}
\setkeys{acs}{maxauthors=10,etalmode=truncate,chaptertitle =true,articletitle=true}

\usepackage{amssymb}
\usepackage{amsmath}
\usepackage[usenames,dvipsnames]{xcolor}
\usepackage{graphicx}
\usepackage{float}
\usepackage[normalem]{ulem}
\usepackage{caption}
\usepackage{subcaption}
\usepackage{natbib}

\newcommand{\ket}[1]{\left|#1\right\rangle}
\newcommand{\braket}[2]{\left\langle#1 |  #2\right\rangle}

\usepackage{xcolor}

\usepackage{multirow}
\usepackage{bm}
\usepackage[version=4]{mhchem} 

\title{Computing X-Ray Absorption Spectra from Linear-Response Particles atop Optimized Holes}

\author{Diptarka Hait}
\email{diptarka@berkeley.edu}
\affiliation
{{Kenneth S. Pitzer Center for Theoretical Chemistry, Department of Chemistry, University of California, Berkeley, California 94720, USA}}
\alsoaffiliation{Chemical Sciences Division, Lawrence Berkeley National Laboratory, Berkeley, California 94720, USA}
\author{Katherine J. Oosterbaan}
\affiliation
{{Kenneth S. Pitzer Center for Theoretical Chemistry, Department of Chemistry, University of California, Berkeley, California 94720, USA}}
\author{Kevin Carter-Fenk}
\author{Martin Head-Gordon}
\email{mhg@cchem.berkeley.edu}
\affiliation
{{Kenneth S. Pitzer Center for Theoretical Chemistry, Department of Chemistry, University of California, Berkeley, California 94720, USA}}
\alsoaffiliation{Chemical Sciences Division, Lawrence Berkeley National Laboratory, Berkeley, California 94720, USA}

\begin{document}

	\maketitle

\begin{abstract}
State specific orbital optimized density functional theory (OO-DFT) methods like restricted open-shell Kohn-Sham (ROKS) can attain semiquantitative accuracy for predicting X-ray absorption spectra of closed-shell molecules. OO-DFT methods however require that each state be individually optimized. In this work, we present an approach to generate an approximate core-excited state density for use with the ROKS energy ansatz, that is capable of giving reasonable accuracy without requiring state-specific optimization. This is achieved by fully optimizing the core-hole through the core-ionized state, followed by use of electron-addition configuration interaction singles (EA-CIS) to obtain the particle level. This hybrid approach can be viewed as a DFT generalization of the static-exchange (STEX) method, and can attain $\sim 0.6$ eV RMS error for the K-edges of C-F through the use of local functionals like PBE and OLYP. This ROKS(STEX) approach can also be used to identify important transitions for full OO ROKS treatment, and can thus help reduce the computational cost for obtaining OO-DFT quality spectra. ROKS(STEX) therefore appears to be a useful technique for efficient prediction of X-ray absorption spectra. 
\end{abstract}

Linear-response time dependent density functional theory\cite{casida1995time,dreuw2005single} (LR-TDDFT, henceforth simply referred to as TDDFT) is currently the most popular method for modeling electronic excited states. The popularity of TDDFT is a consequence of several factors, with perhaps the most important ones being computational efficiency and ability to simultaneously calculate multiple states. These two factors permit TDDFT to appear near `black-box' to casual users, as the lowest $n$ roots for a given system can be usually obtained by requesting software packages to compute $N$ roots (where $N$ is larger than $n$ by a factor of 2-5), for relatively low computational cost and without any prior knowledge about the nature of the states.  TDDFT is also formally exact\cite{runge1984density}, although this is of little practical relevance as the required exact time-dependent exchange-correlation (xc) kernel is unknown and approximate time-independent (ground state) density functionals have to be usually employed\cite{dreuw2005single}. A particularly difficult challenge for TDDFT is that the LR protocol magnifies ground state delocalization errors\cite{perdew1982density,hait2018delocalization} to catastrophic proportions in the excited state.  This has been widely recognized and studied for long-range charge-transfer (CT) states\cite{dreuw2003long,dreuw2004failure,dreuw2005single}, with the TDDFT excitation energies being extremely sensitive to the proportion of Hartree-Fock (HF) exchange employed in the xc functional. TDDFT using time-independent kernels (the so-called adiabatic local density approximation or ALDA\cite{dreuw2005single}) is also incapable of describing double excitations\cite{maitra2004double}, leading to poor performance for open-shell systems\cite{li2016critical,hait2020accurate} and in describing bond dissociation\cite{hait2019beyond}. 

Core-level excitations are another well-known regime of TDDFT failure, as the LR protocol is unable to fully describe the effect of forming a core-hole. Excitation energy errors of $\sim$ 10 eV (vs experimental X-ray absorption spectra) are typical for second period elements like C,N,O and F, ranging from systematic underestimation with pure density functionals like BLYP\cite{besley2010time} to systematic overestimation for pure HF\cite{oosterbaan2018non}. Even larger errors are observed for heavier elements. It is however worth noting that some highly specialized functionals have been developed for predicting X-ray absorption energies\cite{besley2009time}, although the overall quality of the predicted spectra often leaves room for improvement\cite{hait2020accurate,hait2021orbital}. Much progress has also been made on wavefunction-based approaches for simulating core spectroscopy\cite{norman2018simulating,vidal2019new,carbone2019analysis,wenzel2014calculating} as the popularity of experimental core spectroscopy grows due to its value as a sensitive site-specific and time-resolved reporter on chemical dynamics.\cite{kraus2018ultrafast,geneaux2019transient}

State-specific orbital optimized (OO) methods\cite{hait2021orbital} offer a more robust route for addressing many of the shortcomings of TDDFT, as they permit relaxation of the excited state density beyond linear response. In particular, they have been quite effective at predicting core-level spectra of second period elements, with the $\Delta$SCF method\cite{gilbert2008self} attaining $\sim$ 0.5 eV errors for singlet K-edge excitation energies of second period elements with the B3LYP\cite{b3lyp} functional\cite{besley2009self}, albeit via employing spin-contaminated Slater determinants intermediate between singlet and triplet. More recently, some of us have utilized the spin-pure restricted open-shell Kohn-Sham (ROKS) method\cite{frank1998molecular,kowalczyk2013excitation} to predict core-level absorption energies of closed-shell molecules\cite{hait2020highly}, and found $\sim$ 0.2 eV root mean squared error (RMSE) with the SCAN functional\cite{SCAN}. Similar performance has also been recently reported for heavier elements\cite{cunha2021relativistic} through the use of the X2C relativistic Hamiltonian\cite{saue2011primer}. OO-DFT methods like ROKS therefore represent an efficient and reliable route for computing core-level absorption spectra. 

Nonetheless, the OO-DFT methods have some well-known drawbacks in comparison to TDDFT. It was historically difficult to converge core-excited states without the solver instead collapsing back down to the ground state\cite{gilbert2008self}. Recent advances in excited state orbital optimization however offer many routes to avoiding this `variational collapse' problem\cite{gilbert2008self,barca2018simple,shea2020generalized,hait2020excited,carter2020state, levi2020variational,ye2017sigma,corzo2022using}, including the square gradient minimization (SGM) method reported by some of us\cite{hait2020excited}. In addition, the state specificity of such methods mean that either prior knowledge is necessary to identify desired states or a large number of candidate states have to be individually optimized, making the process considerably more computationally demanding \textit{despite} favorable scaling with system size\cite{hait2021orbital}. In particular, many iterations are often spent trying to converge energies to $10^{-6}$ hartrees or lower, which seems quite redundant for core-excitations where uncertainties on the scale of $10^{-3}$ hartrees (0.03 eV) are unlikely to make an enormous difference as experimental uncertainties are typically of the $\sim$ 0.1 eV scale. However, this does not mean that looser convergence thresholds should be routinely employed for core-level calculations, as it would compromise the precise reproducibility of results (while preserving qualitative agreement) and likely lower the quality of other excited state properties like oscillator strengths or forces. 

It is therefore desirable to have protocols that can predict core-excitation energies and properties in a well-defined, reproducible manner, \textit{without} too many excited state specific iterations. There are hybrid schemes that combine OO core-hole relaxation with LR excited state computations, with the best known example being static exchange (STEX)\cite{aagren1994direct,aagren1997direct}. The STEX approach converges core-ionized states with $\Delta$SCF/HF, and then performs configuration interaction singles\cite{foresman1992toward} (CIS) with the resulting orbitals. Finally, any contribution from the RHF ground state in the final wavefunction removed via projection, although this last contribution is generally quite small as the two states are quite different in character. Ignoring this projection correction, STEX can thus be seen as electron-addition CIS (EA-CIS) atop a core-ionized reference, and is thus LR based (CIS being equivalent to TDHF within the Tamm-Dancoff approximation\cite{hirata1999time}). STEX was subsequently generalized to nonorthogonal CIS\cite{oosterbaan2018non,oosterbaan2019non,oosterbaan2020generalized} (NOCIS) for systems with multiple symmetry equivalent atoms. However, use of HF leads to $\sim$ 1.4 eV overestimation in core-level excitation energies of closed-shell molecules with STEX/NOCIS (as can be seen from Table \ref{tab:lightKedges}), especially for core$\to$valence excitations (the errors being generally lower for core$\to$Rydberg processes). This is perceptibly greater than the $\sim 0.6$ eV error with fully OO ROKS/HF\cite{hait2021orbital}. This difference also leads to an impression that STEX spectrum are sometimes `compressed' relative to experiment\cite{norman2018simulating}, as the low lying states with larger valence character are blueshifted much more than the higher energy Rydberg states. 


The difference between STEX and ROKS/HF can be interpreted to 
arise from ROKS/HF fully optimizing the density in the presence of both the particle and the hole, while STEX only optimizes the density in the presence of the hole and then determines the effect of the particle via CIS. The latter is equivalent to CIS estimating the particle level through a linear combination of virtual orbitals of the core-ionized state. This can be mathematically demonstrated as follows. Let the ROHF determinant for the core ionized state be $\ket{\Phi_+}$, the core-hole orbital be $h$ and $\{a,b,c\ldots\}$ the unoccupied orbitals (with there being $V$ such orbitals). Let us further define $\ket{\Phi'}=a_h^\dagger \ket{\Phi_+}$ as the neutral, RHF, core-hole filled state with ROHF core-ionized orbitals. CIS for singlet states therefore involves diagonalization of the Hamiltonian within the subspace spanned by $\ket{\Phi {}^a_h}=\dfrac{1}{\sqrt{2}}\left(a^\dagger_a a_h+a^\dagger_{\bar{a}} a_{\bar{h}}\right)\ket{\Phi'},\ket{\Phi {}^b_h}=\dfrac{1}{\sqrt{2}}\left(a^\dagger_b a_h+a^\dagger_{\bar{b}} a_{\bar{h}}\right)\ket{\Phi'}$ etc. 
The matrix representation of the Hamiltonian within this $V$ dimensional subspace of singly excited singlet states is given by:
\begin{align}
    A_{ab}&=E'\delta_{ab}+F'_{ab}-F'_{hh}\delta_{ab}+2\braket{hh}{ab}-\braket{ha}{hb}
\end{align}
where the Fock operator $F'$ is constructed from $\ket{\Phi'}$ (and not $\ket{\Phi_+}$), while $E'$ is the energy of  $\ket{\Phi'}$. If we solve the $\mathbf{AX}=E\mathbf{X}$ CIS eigenproblem, the eigenstates are given by:
\begin{align}
    \ket{\Phi_h^p}&=\dfrac{1}{\sqrt{2}}\displaystyle\sum\limits_{a} X_{pa}\left(a^\dagger_a a_h+a^\dagger_{\bar{a}} a_{\bar{h}}\right)\ket{\Phi'}=\left(a^\dagger_p a_h+a^\dagger_{\bar{p}} a_{\bar{h}}\right)\ket{\Phi'}\\
    a^\dagger_p&=\displaystyle\sum\limits_{a} X_{pa} a^\dagger_a \label{eqn:peqn}
\end{align}
The final result is therefore a $h\to p$ excitation from $\ket{\Phi'}$, with the particle level $p$ being a linear combination of the original unoccupied levels $\{a\}$ that is obtained via the CIS procedure through Eq. \ref{eqn:peqn}. This connection implies that each unprojected STEX eigenstate has the form of a spin-adapted open-shell HF wavefunction, analogous to ROKS. However, the hole orbital $h$ and other doubly occupied levels are found via optimization of the core-ionized state $ \ket{\Phi_+}$, in contrast to the fully OO ROKS/HF approach that is specific to each excited singlet state. 

STEX's use of an unoptimized, CIS particle level atop core-ionized orbitals instead of a fully optimized ROKS/HF density should lead to an overestimation of excitation energies, at least for the lowest energy core-excited states. 
This appears to suggest that a state-specific DFT approach that utilizes the spin-adapted STEX wavefunction $\ket{\Phi_h^p}$ would need to take advantage of some error cancellation in order to have low error. In particular, several generalized gradient approximations (GGAs) like PBE\cite{PBE} and B97-D\cite{b97d} systematically underestimate core-ionization energies by $\sim 1$ eV with $\Delta$SCF, suggesting that their use in STEX like protocols can lead to low net error via cancellation of the functional specific underestimation for core-hole formation with the overestimation arising from use of CIS particle levels.  The error cancellation in such models can be compared to the manner in which hybrid functionals attempt to mitigate the delocalization error in local KS approaches with the overlocalizing tendency of HF\cite{b3lyp,hait2018delocalization}. Such a protocol also has computational advantages, as the OO procedure is necessary only for ionizing every relevant core orbital $h$, instead of having to independently optimize a particle-hole pair for each excited state with ROKS. In addition, use of pure GGAs avoids the computational burden of computing exact exchange (aside from the single construction of $\mathbf{A}$ per site).

We therefore propose a scheme for computing core-level excitation energies using electronic configurations generated via a STEX like protocol, but using the state-specific ROKS energy ansatz. This requires identification of the particle and hole levels corresponding to the excitation, which we determine as follows:
\begin{enumerate}
    \item Converge restricted KS equations for the ground state to obtain energy $E_0$. 
    \item Converge the RO core-ionized state with the same functional. 
    \item Perform CIS atop $\ket{\Phi'}$ generated from the core-ionized RO orbitals and find the particle levels via Eq. \ref{eqn:peqn}. Note that this step explicitly uses HF, irrespective of the KS functional used in the preceding steps (i.e. $F'$ etc are found from HF, acting upon KS orbitals).
    \item For a given particle level $p$ obtained in the previous step, find the excited state singlet ROKS energy $E_S$. Let $E_M$ be the KS energy of the spin-contaminated determinant $a^\dagger_p a_i\ket{\Phi'}$ and $E_T$ the energy of the pure triplet $a^\dagger_p a_{\bar{i}}\ket{\Phi'}$. The determinant $a^\dagger_p a_i\ket{\Phi'}$ is half-singlet and half-triplet\cite{frank1998molecular,ziegler1977calculation}, and so $E_S=2E_M-E_T$.
    \item The excitation energy is then $E_S-E_0$.
\end{enumerate}
The use of HF in step 3 irrespective of the chosen KS functional is intentional, in order to have a wavefunction theory based definition of the particle level and to avoid use of KS eigenvalues anywhere in the problem. The use of HF also allows for direct inclusion of relevant double excitations in open-shell systems via XCIS\cite{maurice1996nature} and related methods\cite{oosterbaan2019non,oosterbaan2020generalized}, which would be less straightforward with a KS treatment. We also note that the eigenvalues of $\mathbf{A}$ (i.e. CIS state energies) are discarded. The only purpose of CIS is to generate a particle level which can be used to obtain ROKS energies. The method can therefore be described as semi-OO, wherein the doubly occupied levels and the core-hole are identical to the fully optimized core-ionized state while the particle level is found from CIS. We subsequently refer to this hybrid method as ROKS(STEX), as opposed to fully orbital optimized ROKS which is simply referred to as ROKS. We note that ROKS(STEX) with HF would be identical to STEX, if the ground state contribution was not projected out in the latter. It can therefore be viewed as a KS based generalization of STEX, as well as an approximation to the usual ROKS protocol.

\begin{table}[htb!]
\footnotesize{
\begin{tabular}{lrrrrrrrrr}
Species   & \multicolumn{1}{l}{Expt} & \multicolumn{1}{l}{SCAN (OO)} & \multicolumn{1}{l}{SCAN} & \multicolumn{1}{l}{PBE} & \multicolumn{1}{l}{PBE0} & \multicolumn{1}{l}{B88\cite{b88}} & \multicolumn{1}{l}{OLYP\cite{olyp,lyp}} & \multicolumn{1}{l}{B97-D} & \multicolumn{1}{l}{STEX} \\
\ce{\textbf{C}2H4}      & 284.7\cite{hitchcock1977carbon}                    & 284.7                         & 285.8                    & 285.0                   & 285.4                    & 285.3                   & 285.2                    & 285.3                     & 286.4                    \\
\ce{H\textbf{C}HO}      & 285.6\cite{remmers1992high}                    & 285.8                         & 287.4                    & 286.5                   & 286.8                    & 286.9                   & 286.6                    & 286.6                     & 288.2                    \\
\ce{\textbf{C}2H2}      & 285.9\cite{hitchcock1977carbon}                    & 285.7                         & 286.5                    & 285.6                   & 286.0                    & 285.9                   & 285.8                    & 285.9                     & 287.3                    \\
\ce{\textbf{C}2N2}      & 286.3\cite{hitchcock1979inner}                    & 286.3                         & 287.3                    & 286.3                   & 286.7                    & 286.7                   & 286.4                    & 286.5                     & 288.3                    \\
\ce{H\textbf{C}N}       & 286.4\cite{hitchcock1979inner}                     & 286.4                         & 287.5                    & 286.6                   & 286.9                    & 286.9                   & 286.7                    & 286.7                     & 288.2                    \\
\ce{Me2\textbf{C}O}     & 286.4\cite{prince2003near}                    & 286.5                         & 288.2                    & 287.3                   & 287.7                    & 287.7                   & 287.4                    & 287.3                     & 289.2                    \\
\ce{\textbf{C}2H6}      & 286.9\cite{hitchcock1977carbon}                    & 286.8                         & 287.2                    & 286.2                   & 286.6                    & 286.0                   & 286.5                    & 286.8                     & 287.5                    \\
\ce{\textbf{C}O}        & 287.4\cite{domke1990carbon}                    & 287.1                         & 288.2                    & 287.3                   & 287.7                    & 287.6                   & 287.3                    & 287.4                     & 289.2                    \\
\ce{\textbf{C}H4}       & 288.0\cite{schirmer1993k}                    & 288.0                         & 288.2                    & 287.1                   & 287.6                    & 286.9                   & 287.4                    & 287.5                     & 288.5                    \\
\ce{\textbf{C}H3OH}      & 288.0\cite{prince2003near}                    & 288.2                         & 288.6                    & 287.5                   & 288.0                    & 287.4                   & 287.8                    & 288.2                     & 289.1                    \\
\ce{H\textbf{C}OOH}     & 288.1\cite{prince2003near}                    & 288.0                         & 289.5                    & 288.5                   & 288.9                    & 288.9                   & 288.6                    & 288.6                     & 290.5                    \\
\ce{H\textbf{C}OF}      & 288.2\cite{robin1988fluorination}                    & 288.2                         & 289.7                    & 288.7                   & 289.1                    & 289.1                   & 288.8                    & 288.8                     & 290.8                    \\
\ce{\textbf{C}O2}       & 290.8\cite{prince1999vibrational}                    & 290.4                         & 291.5                    & 290.3                   & 290.9                    & 290.8                   & 290.4                    & 290.4                     & 293.1                    \\
\ce{\textbf{C}F2O}      & 290.9\cite{robin1988fluorination}                    & 290.6                         & 292.1                    & 291.0                   & 291.6                    & 291.5                   & 291.1                    & 291.2                     & 293.6                    \\
\ce{C2\textbf{N}2}      & 398.9\cite{hitchcock1979inner}                     & 398.9                         & 399.7                    & 398.8                   & 399.1                    & 399.2                   & 398.8                    & 398.6                     & 400.4                    \\
\ce{HC\textbf{N}}       & 399.7\cite{hitchcock1979inner}                     & 399.7                         & 400.5                    & 399.6                   & 399.9                    & 399.9                   & 399.6                    & 399.5                     & 401.0                    \\
Imidazole (N) & 399.9\cite{apen1993experimental}                    & 399.9                         & 400.9                    & 400.0                   & 400.4                    & 400.2                   & 400.0                    & 399.9                     & 401.3                    \\
\ce{\textbf{N}H3}       & 400.8\cite{schirmer1993k}                    & 400.5                         & 401.0                    & 399.9                   & 400.3                    & 399.8                   & 400.1                    & 400.2                     & 401.2                    \\
\ce{\textbf{N}2}        & 400.9\cite{myhre2018theoretical}                    & 400.9                         & 402.1                    & 401.2                   & 401.5                    & 401.5                   & 401.1                    & 400.9                     & 402.6                    \\
\ce{\textbf{N}NO}       & 401.0\cite{prince1999vibrational}                    & 401.1                         & 402.4                    & 401.4                   & 401.7                    & 401.8                   & 401.4                    & 401.2                     & 402.8                    \\
Glycine (N)   & 401.2\cite{plekan2007x}                    & 401.1                         & 401.7                    & 400.7                   & 401.1                    & 400.4                   & 400.8                    & 400.9                     & 402.0                    \\
Pyrrole (N)   & 402.3\cite{pavlychev1995nitrogen}                    & 402.3                         & 403.0                    & 402.0                   & 402.4                    & 401.6                   & 402.1                    & 401.9                     & 402.9                    \\
Imidazole (NH) & 402.3\cite{apen1993experimental}                    & 402.4                         & 403.1                    & 402.2                   & 402.6                    & 401.9                   & 402.2                    & 402.1                     & 403.3                    \\
\ce{N\textbf{N}O}       & 404.6\cite{prince1999vibrational}                    & 404.5                         & 405.7                    & 404.7                   & 405.1                    & 405.0                   & 404.5                    & 404.4                     & 406.6                    \\
\ce{HCH\textbf{O}}      & 530.8\cite{remmers1992high}                     & 530.9                         & 532.1                    & 531.3                   & 531.4                    & 531.6                   & 531.0                    & 530.7                     & 531.9                    \\
\ce{Me2C\textbf{O}}     & 531.4\cite{prince2003near}                    & 531.3                         & 532.7                    & 531.8                   & 531.9                    & 532.2                   & 531.5                    & 531.1                     & 532.8                    \\
\ce{HC\textbf{O}F}      & 532.1\cite{robin1988fluorination}                    & 532.1                         & 533.3                    & 532.4                   & 532.5                    & 532.7                   & 532.2                    & 531.8                     & 533.2                    \\
\ce{HCOOH} (O)     & 532.2\cite{prince2003near}                    & 532.0                         & 533.2                    & 532.3                   & 532.4                    & 532.6                   & 532.0                    & 531.6                     & 533.1                    \\
\ce{CF2\textbf{O}}      & 532.7\cite{robin1988fluorination}                    & 533.1                         & 534.4                    & 533.5                   & 533.6                    & 533.9                   & 533.3                    & 532.8                     & 534.4                    \\
\ce{H2\textbf{O}}       & 534.0\cite{schirmer1993k}                    & 533.9                         & 534.4                    & 533.2                   & 533.6                    & 533.2                   & 533.2                    & 533.0                     & 534.5                    \\
\ce{CH3\textbf{O}H}      & 534.1\cite{prince2003near}                    & 534.0                         & 534.7                    & 533.6                   & 533.9                    & 533.4                   & 533.5                    & 533.3                     & 534.5                    \\
\ce{C\textbf{O}}        & 534.2\cite{domke1990carbon}                    & 534.2                         & 535.2                    & 534.3                   & 534.4                    & 534.5                   & 534.0                    & 533.6                     & 534.9                    \\
\ce{NN\textbf{O}}       & 534.6\cite{prince1999vibrational}                    & 535.1                         & 536.3                    & 535.4                   & 535.3                    & 535.7                   & 535.2                    & 534.8                     & 535.7                    \\
Furan  (O)    & 535.2\cite{duflot2003core}                    & 535.2                         & 535.9                    & 534.9                   & 535.2                    & 535.1                   & 534.8                    & 534.3                     & 536.0                    \\
\ce{HCOOH} (OH)     & 535.4\cite{prince2003near}                    & 535.5                         & 536.5                    & 535.5                   & 535.8                    & 535.8                   & 535.4                    & 535.0                     & 536.5                    \\
\ce{C\textbf{O}2}       & 535.4\cite{prince1999vibrational}                    & 535.6                         & 536.6                    & 535.5                   & 535.7                    & 535.9                   & 535.3                    & 534.9                     & 536.7                    \\
\ce{\textbf{F}2}        & 682.2\cite{hitchcock1981k}                    & 682.5                         & 685.3                    & 684.4                   & 684.3                    & 685.0                   & 683.8                    & 683.1                     & 684.3                    \\
\ce{H\textbf{F}}        & 687.4\cite{hitchcock1981k}                    & 687.5                         & 688.0                    & 686.7                   & 687.0                    & 686.8                   & 686.4                    & 686.0                     & 687.7                    \\
\ce{HCO\textbf{F}}      & 687.7\cite{robin1988fluorination}                    & 688.0                         & 688.9                    & 687.8                   & 688.0                    & 688.2                   & 687.4                    & 686.9                     & 689.0                    \\
\ce{C\textbf{F}2O}      & 689.2\cite{robin1988fluorination}                    & 689.6                         & 690.8                    & 689.6                   & 689.8                    & 690.0                   & 689.2                    & 688.6                     & 690.8                    \\
          & \multicolumn{1}{l}{}     & \multicolumn{1}{l}{}          & \multicolumn{1}{l}{}     & \multicolumn{1}{l}{}    & \multicolumn{1}{l}{}     & \multicolumn{1}{l}{}    & \multicolumn{1}{l}{}     & \multicolumn{1}{l}{}      & \multicolumn{1}{l}{}     \\
RMSE      & \multicolumn{1}{l}{}     & 0.20                          & 1.18                     & 0.58                    & 0.62                     & 0.82                    & 0.51                     & 0.55                      & 1.56                     \\
ME        & \multicolumn{1}{l}{}     & 0.01                          & 1.05                     & 0.06                    & 0.37                     & 0.29                    & 0.01                     & -0.14                     & 1.40                     \\
MAX       & \multicolumn{1}{l}{}     & 0.49                          & 3.09                     & 2.21                    & 2.07                     & 2.83                    & 1.60                     & 1.39                      & 2.76                    
\end{tabular}
}
\caption{Lowest dipole allowed K-edge excitation energies (in eV) for 40 small molecules, at the C,N,O,F K-edges. The site of the excitation is bolded (or otherwise specified within parentheses). Root mean squared error (RMSE), mean signed error (ME), maximum absolute error (MAX) and mean absolute error (MAE) vs experiment are also reported. Further details about calculations are reported in the computational methods section.}
\label{tab:lightKedges}
\end{table}

It is worthwhile examining the performance of this ROKS(STEX) approach against experiment to determine viability. Table \ref{tab:lightKedges} reports performance with a number of GGA functionals (along with the SCAN meta-GGA and the PBE0\cite{pbe0} hybrid GGA) for the lowest dipole allowed excitation in 40 small molecules. The dataset is dominated by $1s\to\pi^*$ transitions, which are amongst the hardest to model with STEX. Fully orbital optimized ROKS/SCAN values are supplied, for comparison.

It is immediately apparent that the ROKS(STEX) approaches are nowhere close in accuracy to full OO ROKS/SCAN, which has a quite low RMSE of 0.2 eV and virtually no systematic bias. On the other hand, ROKS(STEX)/SCAN overestimates energies by $\sim 1$ eV due to the suboptimal nature of the CIS derived particle orbital as compared to full orbital relaxation. SCAN predicts quite accurate $\Delta$SCF core-electron binding energies\cite{kahk2019accurate,kahk2021core,hait2020highly}, and therefore the error arising from the unrelaxed particle level remains uncancelled. 
ROKS(STEX)/SCAN is nonetheless a perceptible improvement over standard STEX, which overestimates by 1.4 eV. It is nonetheless worth noting that STEX has a lower error for \ce{F2} than ROKS(STEX)/SCAN, resulting in a smaller maximum absolute error. 

However, use of functionals that systematically underestimate core-level excitation energies with ROKS can lead to lower error with ROKS(STEX). This is quite visible for the GGAs OLYP, PBE and B97-D (with the full ROKS numbers being provided in the supporting information). These three GGAs have a reasonably low ROKS(STEX) RMSE of 0.5-0.6 eV, which is nonetheless $3\times$ larger than what can be obtained from ROKS/SCAN. Perhaps more importantly, the maximum error for all three exceeds 1 eV (while ROKS/SCAN only has 0.5 eV maximum deviation from experiment), with \ce{F2} and the C K-edge of HCHO being particularly challenging cases where ROKS(STEX) continues to greatly overestimate.

\begin{figure}[htb!]
    \centering
\begin{minipage}{0.48\textwidth}
    \centering
    \includegraphics[width=\linewidth]{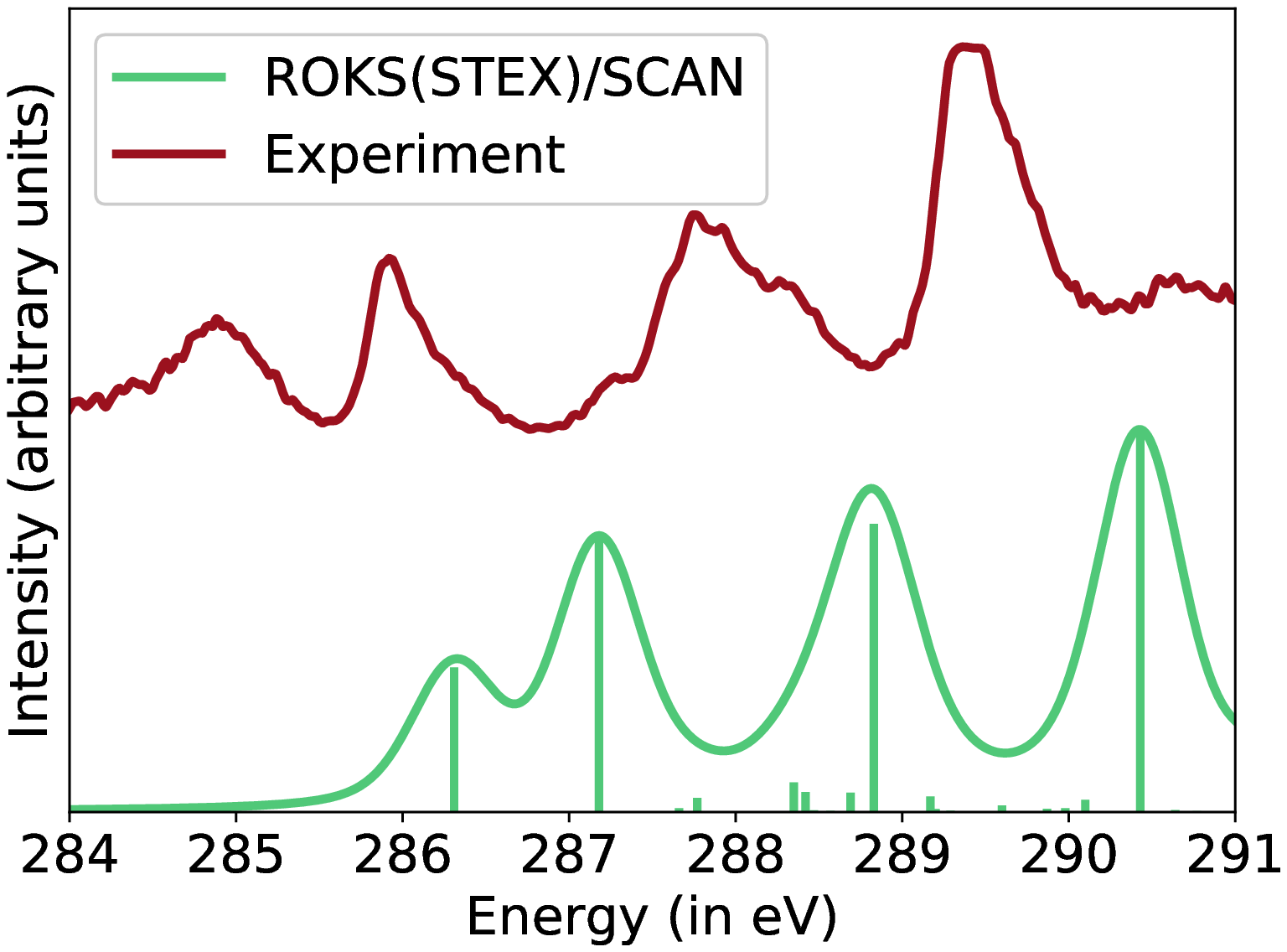}
\end{minipage}\hfill
\begin{minipage}{0.48\textwidth}
    \centering
    \includegraphics[width=\linewidth]{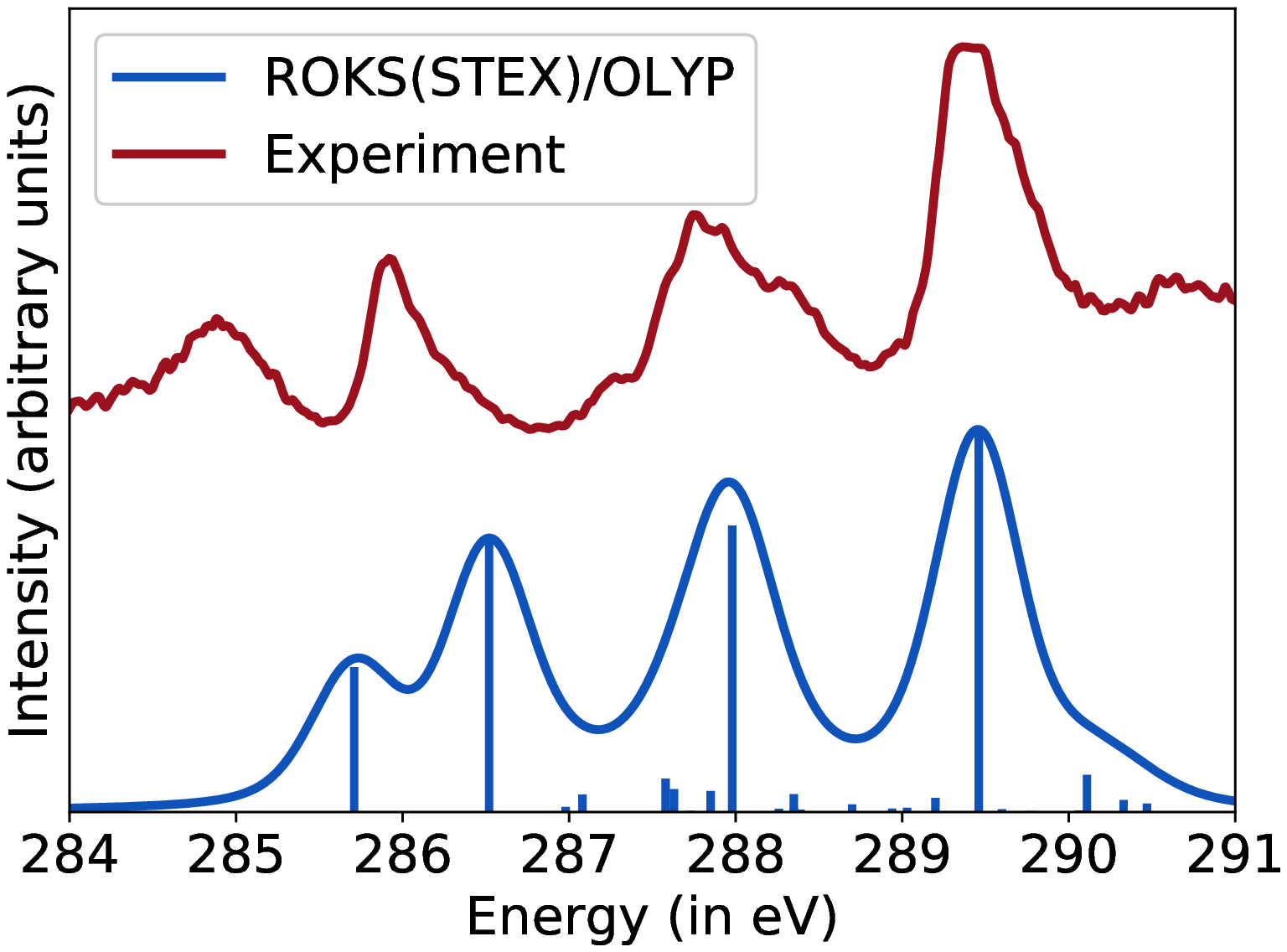}
\end{minipage}
\caption{C K-edge XAS of thymine computed from ROKS(STEX) with SCAN (left) and OLYP (right) as compared to experiment\cite{plekan2008theoretical}. The computed peaks were broadened by a Voigt profile with a Gaussian $\sigma=0.2$ eV and Lorentzian $\gamma = 0.121$ eV.}
\label{fig:thymine}
\end{figure}

We next consider the quality of the complete spectrum predicted by ROKS(STEX). Fig. \ref{fig:thymine} presents the experimental C K-edge of thymine\cite{plekan2008theoretical} vs ROKS(STEX) results from SCAN and OLYP. It can be seen that the general shape of the spectrum is reproduced well, although the peak positions are quite suboptimal. The peaks computed with SCAN are $\sim 1$ eV over experimental results, and reasonable agreement can be found if the computed spectrum is redshifted by 1 eV. On the other hand, the OLYP spectrum is quite compressed, with agreement for the two highest energy peaks being quite adequate, while the lowest energy peak is overestimated by $\sim 0.7$ eV. The compression of the spectrum here is particularly interesting, since all the peaks arise from $1s\to\pi^*$ transitions. Nonetheless, this behavior is consistent with the results reported in Table \ref{tab:lightKedges}, where OLYP overestimated the C K-edge  $1s\to \pi^*$ excitation of HCHO by 1 eV, while being essentially spot on for CO. The ROKS(STEX) excitation energies are therefore nowhere as accurate as full ROKS with OO, leading to deleterious consequences for the overall spectrum. 

On a more optimistic note, oscillator strengths from the ROKS(STEX) approach seem to be adequately accurate. It is thus possible to envision a procedure in which ROKS(STEX) is used to identify the \textit{significant} contributors to the absorption spectrum, followed by full OO on these states to have a more accurate result. Such a protocol thus be effective in reducing computational cost by screening out many weakly absorbing Rydberg type excitations for which full OO is not necessary and ROKS(STEX) energies/oscillator strengths are adequate.  

\begin{figure}[htb!]
    \centering
\begin{minipage}{0.48\textwidth}
    \centering
    \includegraphics[width=\linewidth]{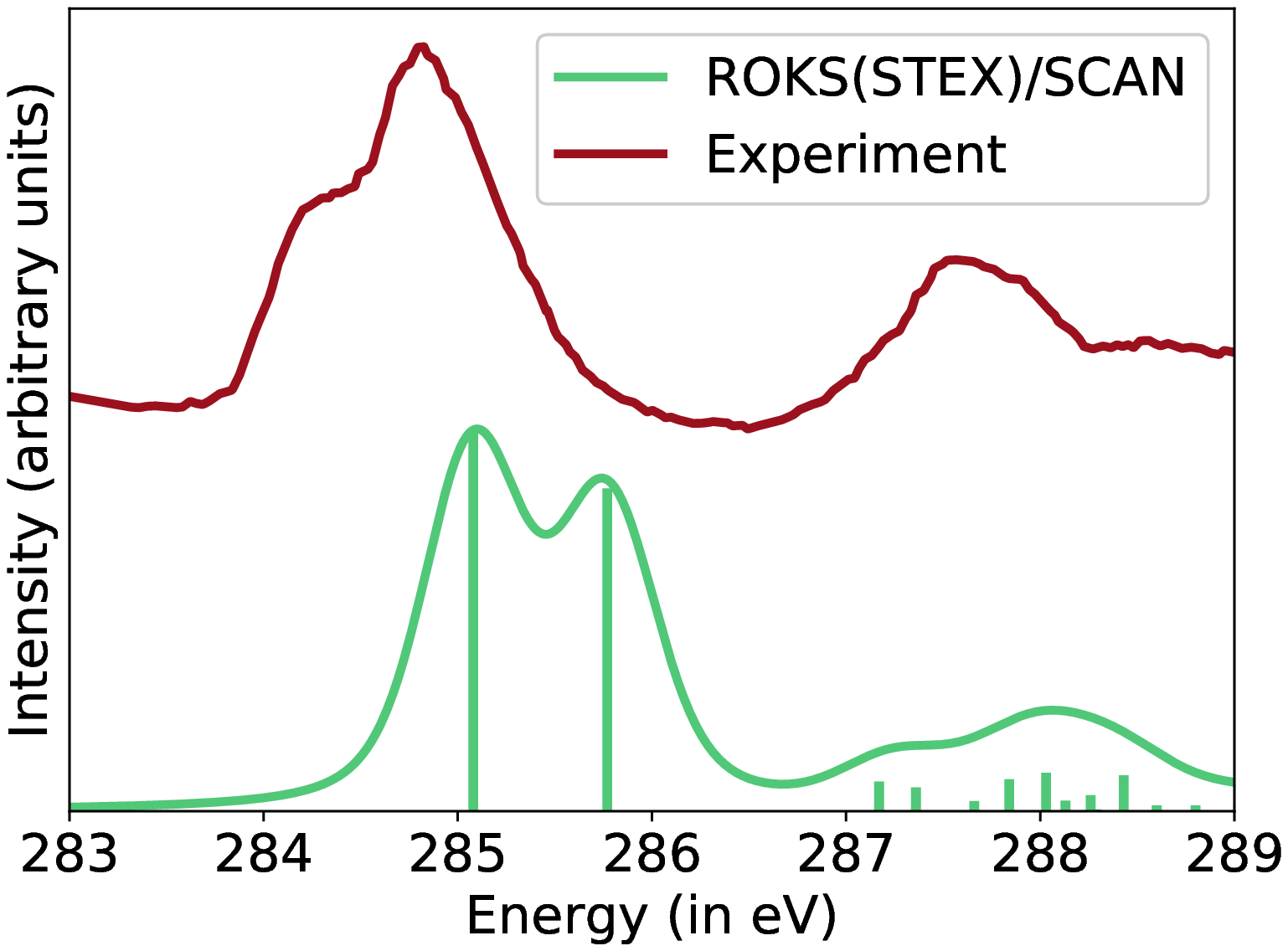}
\end{minipage}\hfill
\begin{minipage}{0.48\textwidth}
    \centering
    \includegraphics[width=\linewidth]{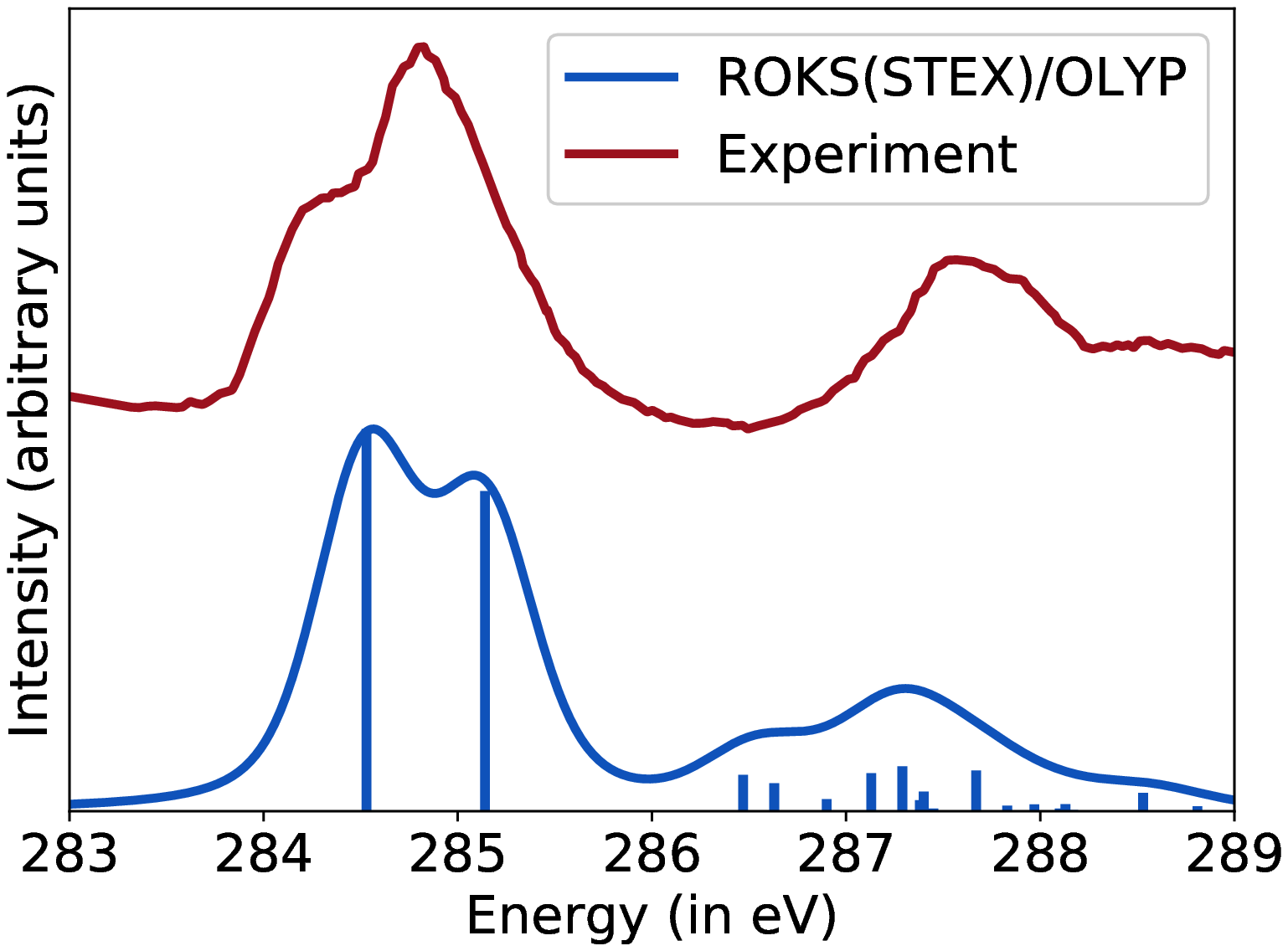}
\end{minipage}
\begin{minipage}{\textwidth}
    \centering
    \subcaption{C K-edge XAS of 1,3-butadiene\cite{sodhi1985high}.}
    \label{fig:butadiene}
\end{minipage}
\begin{minipage}{0.48\textwidth}
    \centering
    \includegraphics[width=\linewidth]{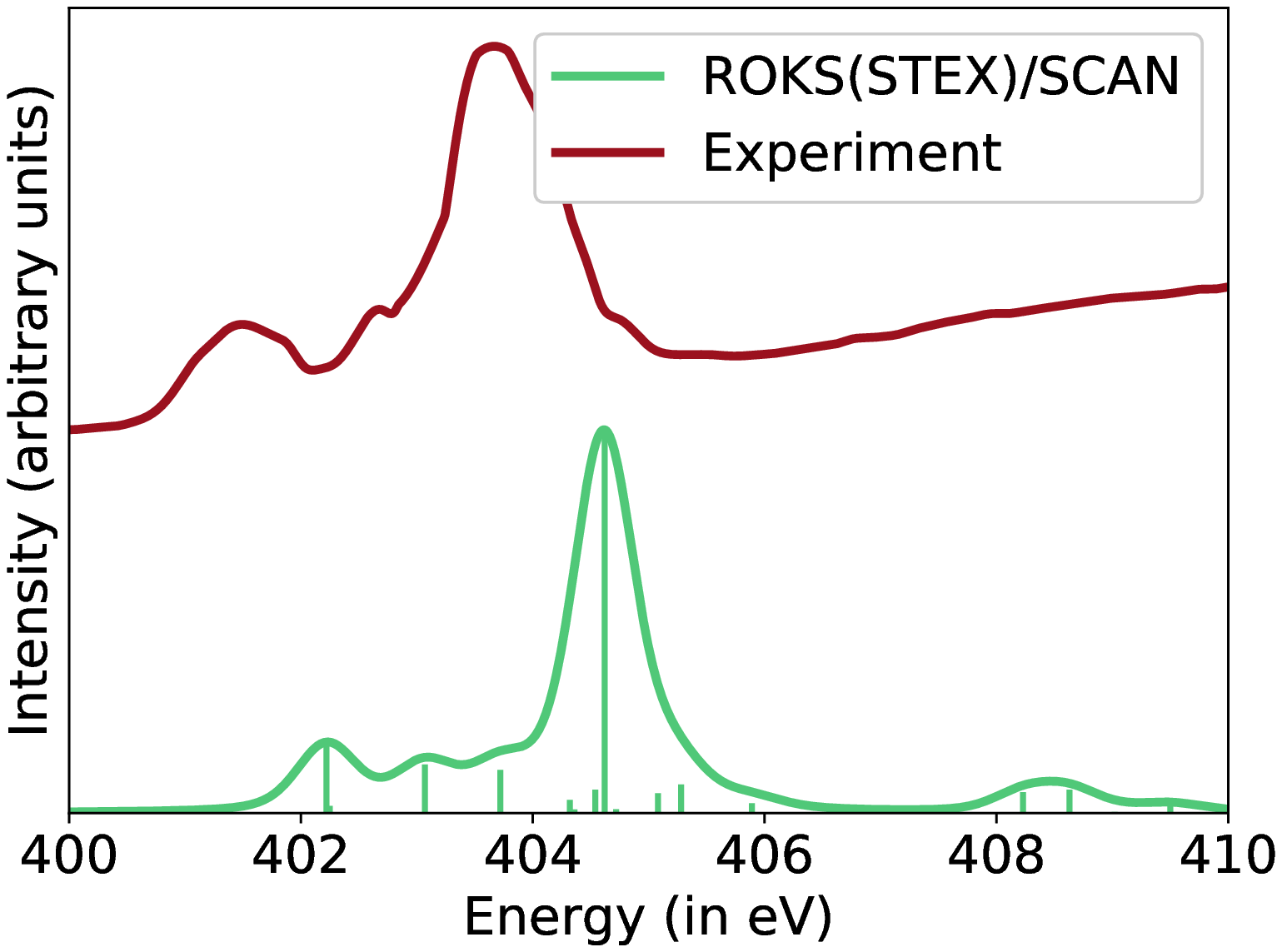}
\end{minipage}\hfill
\begin{minipage}{0.48\textwidth}
    \centering
    \includegraphics[width=\linewidth]{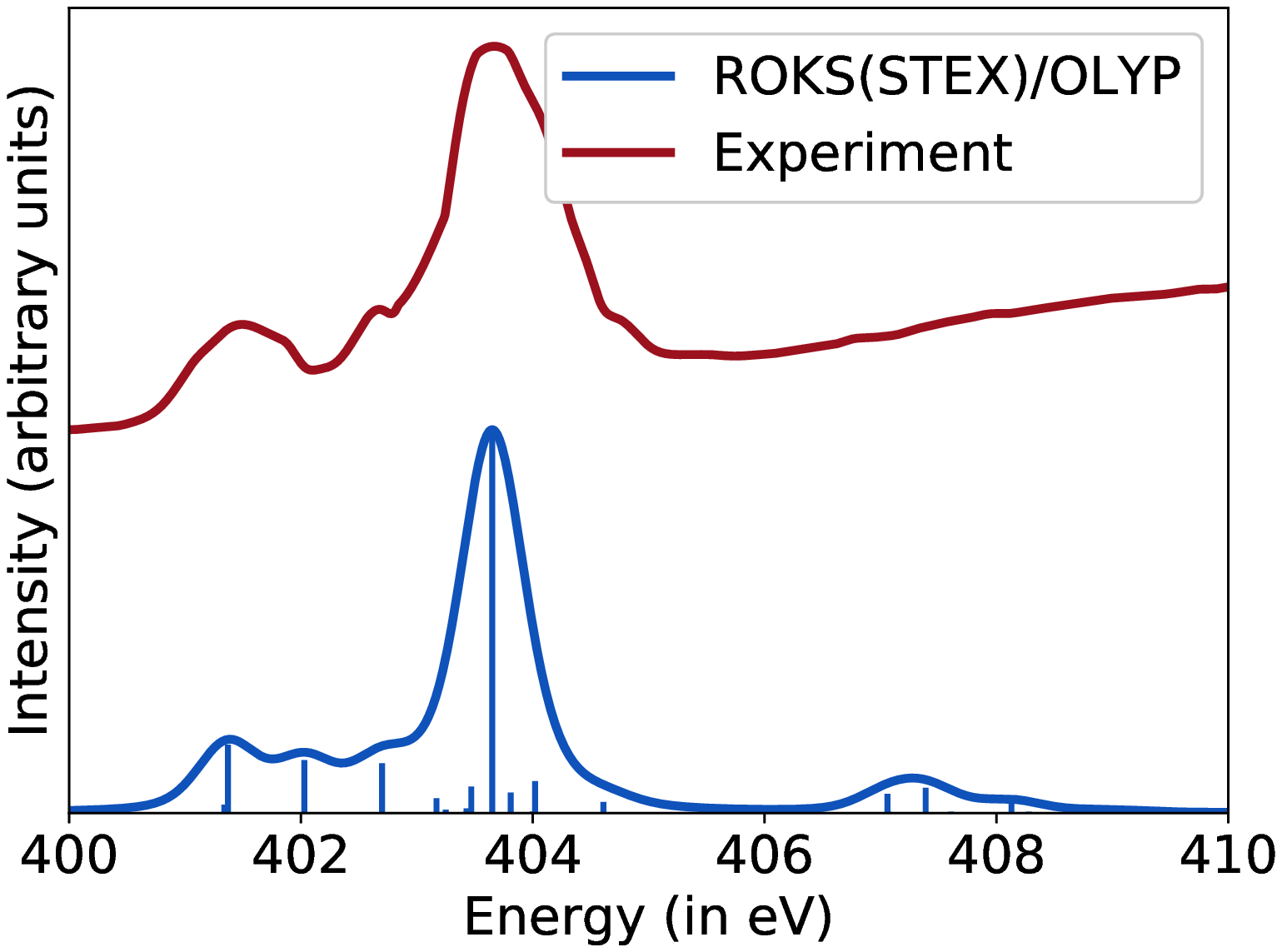}
\end{minipage}
\begin{minipage}{\textwidth}
    \centering
    \subcaption{N K-edge XAS of 4-nitroaniline\cite{turci1996inner}.}
    \label{fig:4nitan}
\end{minipage}
\begin{minipage}{0.48\textwidth}
    \centering
    \includegraphics[width=\linewidth]{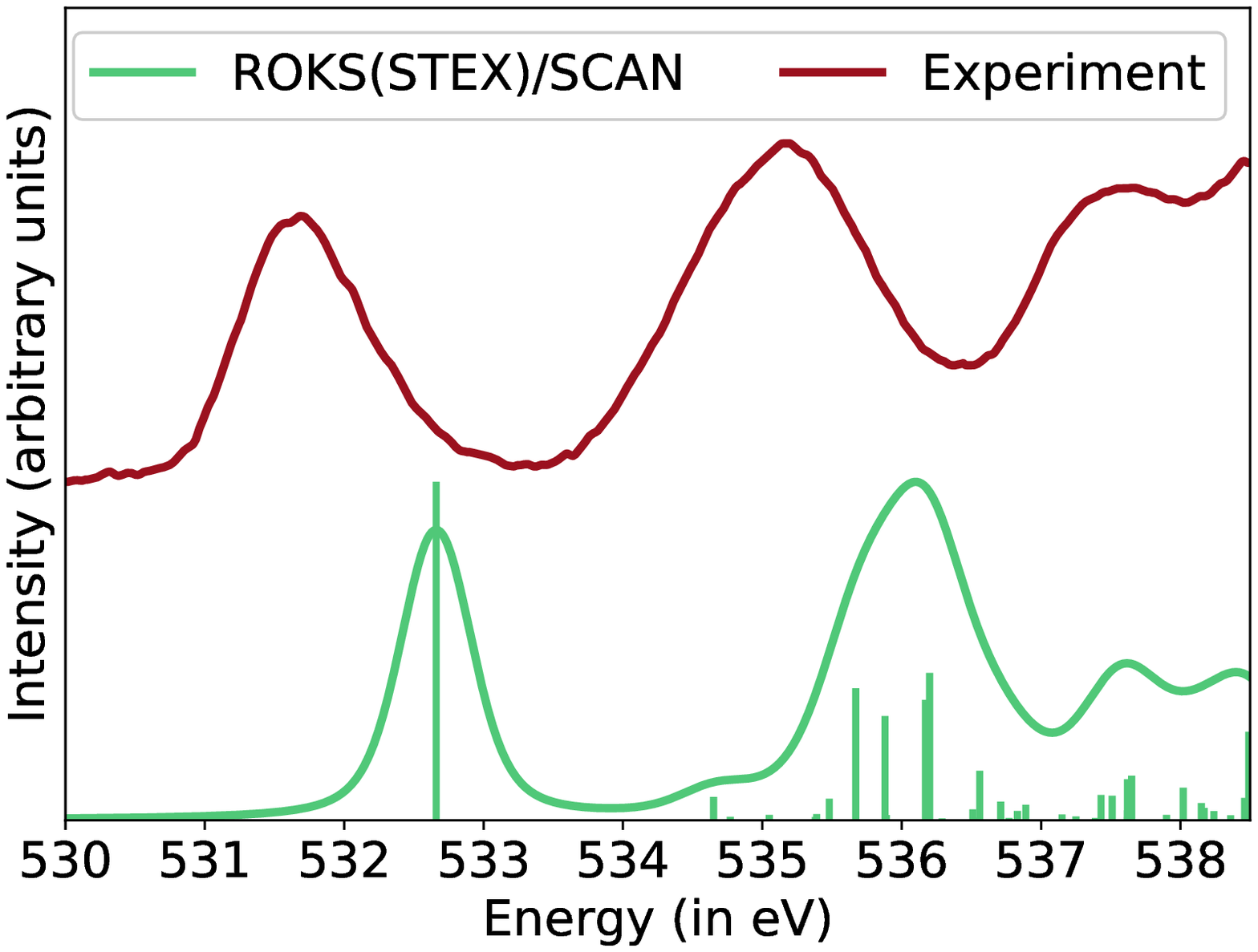}
\end{minipage}\hfill
\begin{minipage}{0.48\textwidth}
    \centering
    \includegraphics[width=\linewidth]{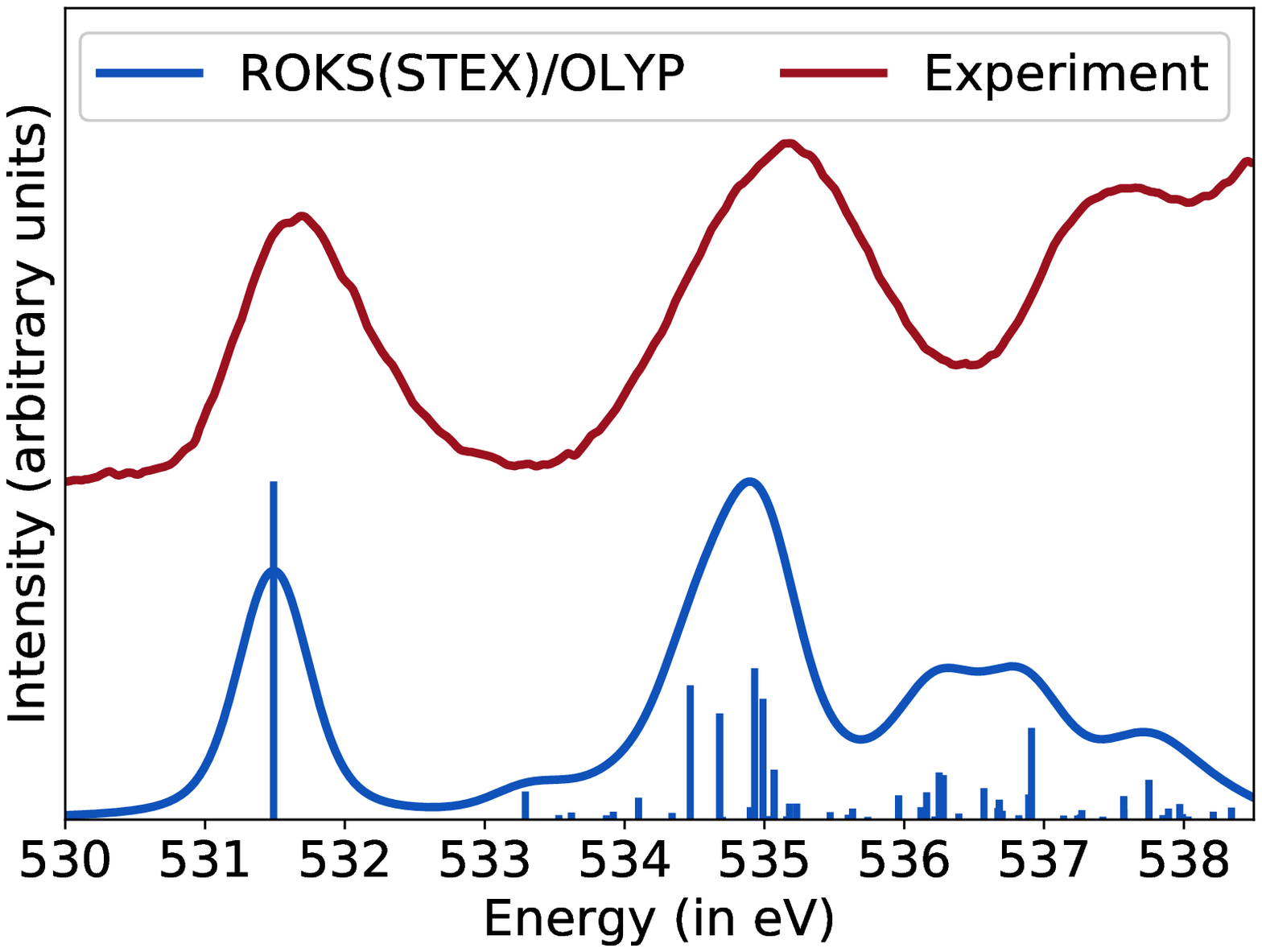}
\end{minipage}
\begin{minipage}{\textwidth}
    \centering
    \subcaption{O K-edge XAS of 4-hydroxybenzoic acid\cite{hill2021positional}.}
    \label{fig:4hydbenz}
\end{minipage}
\caption{ ROKS(STEX) spectra compared to experiment. Computed peaks were broadened by a Voigt profile with a Gaussian $\sigma= 0.2 $ eV and Lorentzian $\gamma = 0.121$ eV.}
\label{fig:good}
\end{figure}

It therefore appears that there are two possible routes to utilize the ROKS(STEX) approach. The first is to use it exclusively for computing spectra with a GGA like OLYP or B97-D, relying on the cancellation of errors between the core-ionization energy underestimation by these functionals and the overestimation from use of the unoptimized STEX particle level. This approach is considerably more computationally efficient than ROKS/SCAN, both because it is core-orbital (site) specific as opposed to state specific, and because it only employs a GGA. The errors nonetheless would be much larger, and the prediction quality can be compromised such as in the case of thymine shown in Fig \ref{fig:thymine}. On the other hand, reasonable results are also possible at times, such as in the case of butadiene, 4-nitroaniline, and 4-hydroxybenzoic acid (as shown in Fig \ref{fig:good}).
\begin{table}[htb!]
\begin{minipage}[b]{0.48\textwidth}
\centering
\begin{tabular}{llll}
State & ROKS   & ROKS(STEX) & Symmetry \\
1     & 533.90 & 534.38      & \ce{A1}     \\
2     & 535.73 & 535.96      & \ce{B2}      \\
3     & 537.04 & 537.07      & \ce{B1}             \\
4     & 537.14 & 537.20      & \ce{A1}             \\
5     & 537.58 & 537.72      &  \ce{A1}                \\
6     & 537.82 & 537.95      & \ce{B2}             \\
7     & 538.24 & 538.26      &    \ce{A1}              \\
8     & 538.25 & 538.28      & \ce{A2}           \\
9     & 538.36 & 538.38      &   \ce{B2}               \\
10    & 538.47 & 538.51      &         \ce{B1}        
\end{tabular}
\subcaption{Excitation energies }
\label{tab:h2ostates}
\end{minipage}\hfill 
\begin{minipage}[b]{0.48\linewidth}
\centering
\includegraphics[width=\columnwidth]{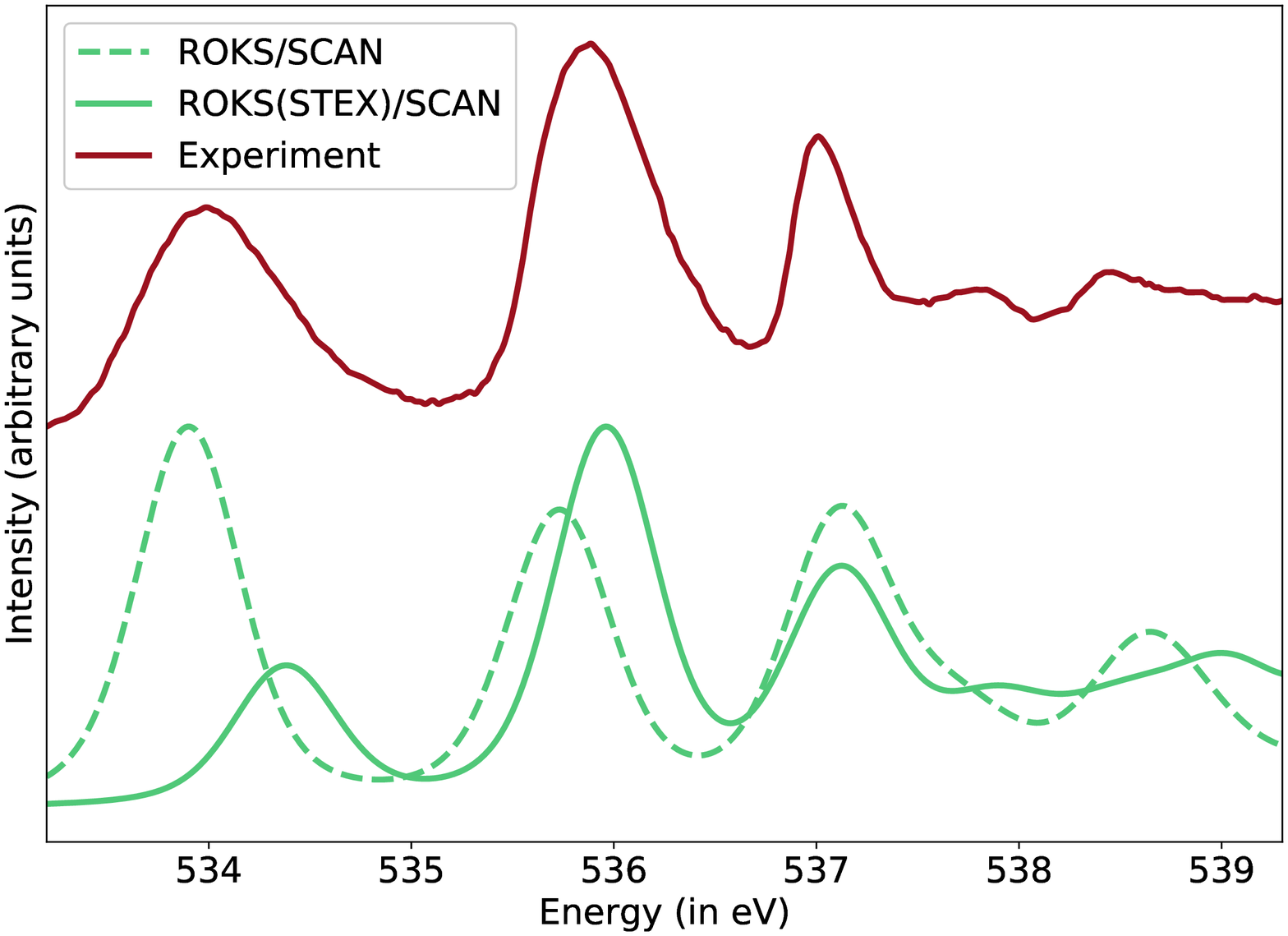}
\subcaption{Computed XAS compared to experiment\cite{schirmer1993k}.}
\label{fig:h2ospec}
\end{minipage}
\caption{O K-edge excitations of \ce{H2O}, via SCAN. Other than the first two states (which are of mixed valence-Rydberg character) the rest are almost purely Rydberg.}\label{tab:h2oall}
\end{table}

The second way to use ROKS(STEX) is in conjunction with full ROKS. ROKS(STEX) with SCAN can be first used to get an approximate sense of how the spectrum will look, and identify the transitions with large oscillator strengths that are below the ionization threshold. These transitions can then be specifically optimized using ROKS, in order to obtain more accurate energies. The remaining low intensity transitions can be left at the ROKS(STEX) level as they will have a rather low effect on the full near edge spectrum. These transitions furthermore will likely contain a large number of Rydberg states, which should be reasonably predicted by the ROKS(STEX) approach ince the particle-hole interaction in such cases will be weak enough to make EA-CIS an acceptable approximation. As evidence, we list the core-excited states of \ce{H2O} and compare the full OO ROKS results with ROKS(STEX) in Table \ref{tab:h2ostates}. It is clear that the Rydberg states predicted by ROKS(STEX) are within 0.1 eV of the full ROKS optimized state, indicating ROKS(STEX)'s efficacy in efficiently modeling such states.  Indeed, the ROKS(STEX) spectrum is very similar in quality to the ROKS spectrum for \ce{H2O}, as shown in Table \ref{tab:h2oall}. 

In conclusion, ROKS(STEX) is a site-specific, computationally efficient method for accessing the core-level spectra of closed-shell molecules. It employs particle levels obtained via EA-CIS atop a core-ionized state, and is therefore less accurate than a fully orbital optimized method like ROKS (with a good functional like SCAN). Nonetheless, cancellation of errors can be carefully employed to have ROKS(STEX) yield results with low systematic error and reasonably low RMSE with GGA functionals like OLYP or B97-D. ROKS(STEX) can also be used to screen excitations a-priori to determine which ones have significant oscillator strengths. These strong intensity transitions can then be accessed via ROKS proper, while the remaining weakly absorbing states are left at the ROKS(STEX) level. 

In future, we intend to investigate the utility of this approach for open-shell systems, where CIS has to be extended to include some double (or higher order) excitations in order to obtain spin-pure results\cite{maurice1996nature}. This can prove useful in interpreting XAS spectra collected to study photochemical dynamics of large systems, as it would assist in decoupling the critical valence excitations that require full OO treatment from the many Rydberg levels that are adequately treated with STEX like approaches.

\section*{Computational methods}
All calculations were performed with a development version of the Q-Chem 5.4 package\cite{epifanovsky2021software}. Local exchange–correlation integrals for DFT were calculated over a radial grid with 99 points and an angular Lebedev grid with 590 points. The spin-free one-electron X2C relativistic Hamiltonian was used for all calculations\cite{cunha2021relativistic,saue2011primer}. For nearly all calculations, the site of the core excitation used an aug-pcX-2 basis\cite{ambroise2018probing}, while aug-pcseg-1\cite{jensen2014unifying} was used for all other atoms. A mixed basis strategy of this nature was previously found to be practically equivalent to purely using the larger basis, for core-level excitation energies.\cite{hait2020highly}
This also served to localize the core-hole onto a single atom for species with equivalent atoms (like O in \ce{CO2}), and thus prevented errors arising from delocalization\cite{perdew1982density,hait2018delocalization} of the hole over multiple sites\cite{hait2020highly}. The one exception regarding basis sets is the data for Table 2, for which the doubly augmented d-aug-pc-2 basis was used instead. The geometries utilized were obtained from Ref \citenum{hait2020highly}. They are also provided in the supporting information, for convenience. 

\section*{Acknowledgment} 
This work was supported by the Director, Office of Science, Office of Basic Energy Sciences, of the U.S. Department of Energy under Contract No. DE-AC02-05CH11231, through the Atomic, Molecular, and Optical Sciences Program of the Chemical Sciences Division of Lawrence Berkeley National Laboratory. Additional support came from the Liquid Sunlight Alliance, which is funded by the U.S. Department of Energy, Office of Science, Office of Basic Energy Sciences, Fuels from Sunlight Hub under Award Number DE-SC0021266. 

\section*{Data Availability}
The data that supports the findings of this study are available within the article and its supplementary material.

\section*{Supporting Information}
XLXS: Raw data.\\
ZIP: Geometries of all species considered in xyz format.
\section*{Conflicts of Interest}
M.H.-G. is a part-owner of Q-Chem, which is the software platform in which the developments described here were implemented.
\bibliography{references}
\end{document}